\documentclass[%
 reprint,
superscriptaddress,
showpacs,
 amsmath,amssymb,
 aps, prl
]{revtex4-1}

\usepackage{graphicx}
\usepackage{dcolumn}
\usepackage{bm}
\usepackage[mathlines]{lineno}
\usepackage{color}

\newcommand{\df}{\; \mathrm{d}}
\newcommand{\eps}{\varepsilon}

\newcommand{\new}[1]{#1}

\begin{document}

\preprint{APS/123-QED}

\title{What controls thermo-osmosis? Molecular simulations show the critical role of interfacial hydrodynamics}

\author{Li Fu}
\affiliation{Univ Lyon, Universit\'e Claude Bernard Lyon 1, CNRS, Institut Lumi\`ere Mati\`ere, F-69622 Villeurbanne, France}
\author{Samy Merabia}%
\affiliation{Univ Lyon, Universit\'e Claude Bernard Lyon 1, CNRS, Institut Lumi\`ere Mati\`ere, F-69622 Villeurbanne, France}
\author{Laurent Joly}%
\email{laurent.joly@univ-lyon1.fr}
\affiliation{Univ Lyon, Universit\'e Claude Bernard Lyon 1, CNRS, Institut Lumi\`ere Mati\`ere, F-69622 Villeurbanne, France}

\date{\today}

\begin{abstract}
Thermo-osmotic and related thermo-phoretic phenomena can be found in many situations from biology to colloid science, but the underlying molecular mechanisms remain largely unexplored. 
Using molecular dynamics simulations, we measured the thermo-osmosis coefficient by both mechano-caloric and thermo-osmotic routes, for different solid-liquid interfacial energies. 
The simulations reveal in particular the crucial role of nanoscale interfacial hydrodynamics. 
For non-wetting surfaces, thermo-osmotic transport is largely amplified by hydrodynamic slip at the interface. For wetting surfaces, the position of the hydrodynamic shear plane plays a key role in determining the \new{amplitude and sign of the} thermo-osmosis coefficient. 
Finally, we measure a giant thermo-osmotic response of the water-graphene interface, which we relate to the very low interfacial friction displayed by this system.
These results open new perspectives for the design of efficient functional interfaces for, e.g., waste heat harvesting. 
\end{abstract}

\maketitle


The fundamental coupling between thermal and hydrodynamic transport in the nanometric vicinity of liquid-solid interfaces has received scanty attention until recently. Thermophoretic phenomena, referring to  the influence of temperature gradients on the flux of colloidal particles, were firstly studied for numerous applications such as optothermal DNA trapping or disease-related protein aggregates identification \cite{Golestanian2007, Dhont2007, Piazza2008, Zambrano2009, Shiomi2009, Wurger:2010dj, Wolff:2016fh, Rajegowda2017}\new{, and this interest for thermophoresis fostered work on its theoretical description~\cite{Demirel2001, Parola2004, Wurger2007, Dhont2008, Burelbach2017}}.
On the other hand, at variance with what has been done for electro-osmosis~\cite{Joly2004, Huang2007, Rotenberg2013, Bonthuis2013, Majumder2015, Maduar2015, Predota2016, Siboulet2017, Bhadauria2017}
and diffusio-osmosis~\cite{Ajdari:2006ci, Marbach2017, Yoshida2017, Lee:2017cw}, very limited theoretical work has been done so far on thermo-osmosis at solid-liquid interfaces. 

Thermo-osmosis was first studied by Derjaguin and Sidorenkov through porous glass \cite{Derjaguin:1941vs} and is usually interpreted as a thermal gradient-induced Marangoni flow~\cite{Ruckenstein:1981kg, Anderson:1989vq, Derjaguin:1987bf}\new{. Advanced continuum descriptions have been developed recently for electrolytes \cite{Dietzel2017}}, but a molecular level understanding is still lacking. \new{In that context}, Ganti et al. \cite{Ganti2017} have explored three different methods to characterize thermo-osmosis using molecular simulations, and they found that all methods yield very similar results. Bregulla et al.~\cite{Bregulla2016} reported the first microscale observation of the velocity field imposed by a nonuniform temperature and deduced the thermo-osmosis coefficient for different surfaces. Nevertheless, the role of surface wettability and interfacial hydrodynamics have hardly been discussed so far.

Here we explore the influence of solid-liquid interfacial energy on thermo-osmosis using molecular dynamics simulations. 
We show the crucial role of interfacial hydrodynamics\new{, explaining in particular the giant thermo-osmosis coefficient observed on non-wetting surfaces and at the water-graphene interface, and controlling the change of sign of the coefficient observed on wetting surfaces.} 

\begin{figure*}
	\centering
	\includegraphics[width=0.8\textwidth]{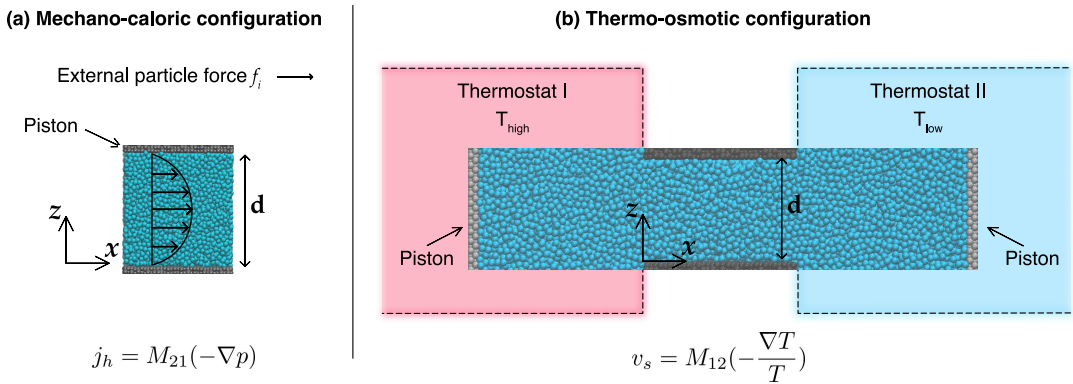}
	\caption{Illustration of the different configurations used to measure the thermo-osmosis coefficient with molecular dynamics simulations.
    (a)~Mechano-caloric route, using a slit nano-channel without reservoirs: a body force per particle $f_i$ is applied to the liquid particles to model an external pressure gradient, and an excess heat flux is generated by the induced Poiseuille flow; (b)~Thermo-osmotic route, where the nanochannel is connected to reservoirs at different temperatures and the thermo-osmotic velocity is measured.}
	\label{fig:Illustration_systems}
\end{figure*}

\paragraph*{Theory}
\label{para:theorie}
The thermo-osmotic response of a fluidic system can be described by the non-diagonal terms of its response matrix \cite{Anderson:1989vq, Derjaguin:1987bf, Kjelstrup:2008iu}:
\begin{equation}
\left[ \begin{array}{c} v_s \\ j_h \end{array} \right] = \begin{bmatrix} M_{11} & M_{12} \\ M_{21} & M_{22} \end{bmatrix} \times \left[ \begin{array}{c} -\nabla p \\ -\nabla T/T \end{array} \right],
\label{eq:coefficient_definition}
\end{equation}
where $v_s$ is the hydrodynamic velocity, $j_h$ is the heat flux density, and $M_{ij}$ are phenomenological coefficients. 
$M_{12}$ is the so-called thermo-osmotic slip coefficient or thermo-osmosis coefficient, which describes the surface-induced flow under a thermal gradient. $M_{21}$ is the so-called mechano-caloric coefficient, which describes the heat flux density generated by a pressure gradient. According to Onsager reciprocal relations, $M_{12} = M_{21}$ \cite{deGroot:1984ue, Brunet:2004bv}. 

Derjaguin and Sidorenkov \new{related the thermo-osmosis coefficient to the interfacial excess enthalpy} using linear non-equilibrium thermodynamics  \cite{Derjaguin:1941vs,Derjaguin:1987bf,Anderson:1989vq}: 
\begin{equation}
M_{12} = M_{21} = \frac{1}{\eta}\int_{0}^{+\infty} z\, \delta h(z) \df z,
\label{eq:M21_linear} 
\end{equation}
where $\eta$ is the liquid viscosity, $z$ the distance to the surface, $z=0$ the position of the interface, $z=+\infty$ the bulk liquid region far from the interface, and $\delta h(z)$ the excess of specific enthalpy as compared to the bulk. 
\new{As detailed in the supplemental material (SM) \footnote{See Supplemental Material at [url] for further details, which includes
Refs. \cite{Jewett:2013fg,Humphrey:1996bk,Bussi2007,Barrat:1999kv,Evans:2016ur,Tocci2014a,Falk2010,KumarKannam2012,Guevara-Carrion2011,Guillaud:2017bk,Markesteijn2012}.}, $\delta h(z)$ can indeed be related to the thermodynamic force acting on the liquid under a thermal gradient, which drives thermo-osmosis.} 
This expression has also been obtained using local thermal equilibrium \cite{Ganti2017} or mechanical \cite{Han:2005ig} routes. 
Note that Eq.~\eqref{eq:M21_linear} is based on a simple, macroscopic view of interfacial hydrodynamics, and it ignores the possible presence of a stagnant liquid layer, or of liquid-solid slip. 
However, for other surface-driven flows, the details of interfacial dynamics can play a key role \cite{Lee:2017cw}. In particular, hydrodynamic slip can amplify electro-osmotic flows \cite{Muller:1986wk, Joly2004, Joly2006, Bouzigues:2008db, Audry:2010gy} and diffusio-osmotic flows \cite{Ajdari:2006ci}, but also thermophoresis of colloids \cite{Morthomas:2009in}. 
Therefore, following previous work in the context of electro-osmosis \cite{Huang2007, Huang2008, BarbosaDeLima2017}, we rewrite Eq.~\eqref{eq:M21_linear}, introducing the shear plane position $z_s$ in order to account for a possible stagnant liquid layer, and the slip length $b$ to describe a possible liquid-solid slip: 
\begin{equation}
M_{12} = M_{21} = \frac{1}{\eta}\int_{z_s}^{+\infty} (z-z_s+b)\, \delta h(z) \df z.
\label{eq:M21_linear_slip} 
\end{equation}
Detailed derivations of the formula presented in this Letter are reported in the SM. 

\begin{figure}
	\centering
	\includegraphics[width=0.35\textwidth]{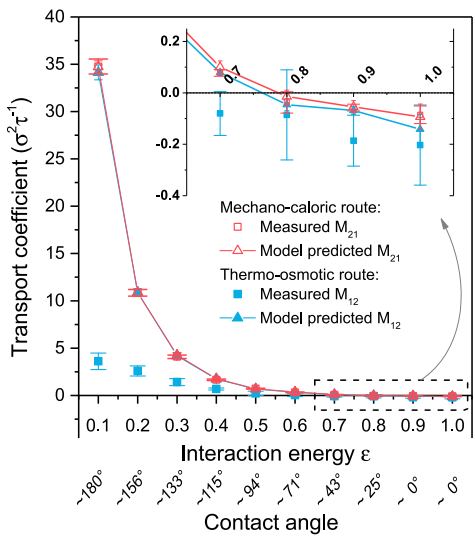}
	\caption{Measured and theoretical thermo-osmosis coefficient for mechano-caloric  and thermo-osmotic configurations, as a function of the solid-liquid interaction energy $\eps_{ls}$ \new{(the corresponding contact angles are indicated below)}. For $\eps_{ls}$ ranging between 0.7 and $1.0\eps$, results are enlarged in the inset. A change of sign is observed for both configurations. A good agreement of predicted and measured coefficient is found for the mechano-caloric configuration, while a large disparity exists for the thermo-osmotic configuration, related to viscous entrance effects (see Fig.~\ref{fig:backflow} and related text).}
	\label{fig:All_M_e}
\end{figure}

\paragraph*{Molecular dynamics simulations}
\label{para:method}
We tested the theoretical prediction with MD simulations, and explored in particular the effect of wetting. All simulations were performed with the LAMMPS package \cite{Plimpton1995}. Technical details can be found in the SM, and here we report only the main features of the models. 
We considered a generic liquid made of particles interacting through a Lennard-Jones (LJ) pair potential, $V (r) = 4 \eps [ ( \sigma /r)^{12} - (\sigma /r) ^6 ]$, with $r$ the distance between the particles, and $\eps$ and $\sigma$ the liquid-liquid interaction energy and distance, respectively. The liquid was confined in a slit configuration between two Einstein solids interacting with the liquid also through a LJ potential
with the same $\sigma$. We varied the liquid-solid wetting properties by adjusting the liquid-solid interaction energy in the range $\eps_\text{ls}= 0.1 \,\eps$ (very hydrophobic) to $1.0  \eps$ (very hydrophilic)\new{; the corresponding contact angles displayed in Fig. \ref{fig:All_M_e} were estimated with sessile droplet simulations, as detailed in the SM}.  

We used both the mechano-caloric and the thermo-osmotic routes to measure the thermo-osmosis coefficient, see Fig.~\ref{fig:Illustration_systems}. 
For the mechano-caloric route (Fig.~\ref{fig:Illustration_systems}a), we considered an infinite slit nanochannel (using periodic boundary conditions). We applied a body force per particle to model a pressure gradient, and measured the resulting heat flux $j_h = \int \delta h(z) v_x(z) \df z$, with $v_x(z)$ the measured velocity profile, and $\delta h(z)$ the excess specific enthalpy profile. 
The local specific enthalpy was expressed as
$h(z)=(u_i(z)+p_i(z))\rho(z)$,
(with $u_i(z)$ 
the energy per particle, $p_i(z)$ the atom-based virial expression for pressure, and $\rho(z)$ the local density, see the SM for detail), and we obtained $\delta h(z)$ by subtracting the bulk value in the middle of the channel. We then computed the mechano-caloric coefficient $M_{21} = j_h / (-\nabla p)$.  
We also calculated a theoretical value for $M_{21}$ according to Eq.~\eqref{eq:M21_linear_slip}, where the hydrodynamic parameters $\eta$, $z_s$ and $b$ were obtained by fitting the numerical velocity profiles. 
For the thermo-osmotic route, 
we connected the slit channel to reservoirs at the same pressure (imposed by two pistons) and different temperatures. We measured the thermo-osmotic velocity $v_s$ from the time evolution of the number of particles in the reservoirs, and the thermal gradient $\nabla T$ and the average temperature $T$ from the linear temperature profile in the channel. We then computed the response coefficient $M_{12} = v_s / (-\nabla T/T)$. Here again we calculated a theoretical value for $M_{12}$ according to Eq.~\eqref{eq:M21_linear_slip}, using the enthalpy profiles measured in the channel.

\paragraph*{Results}
\label{Results}

We first focus on the measurements of $M_{21}$ using the mechano-caloric route. Figure~\ref{fig:All_M_e} presents the evolution of measured and theoretical $M_{21}$ as a function of solid-liquid interaction energy $\eps_{ls}$ (red open symbols). 
Within uncertainties, Eq.~\eqref{eq:M21_linear_slip} predicts well the measured values of $M_{21}$. 
Using typical molecular lengths ($\sigma = 0.34$\,nm) and times ($\tau = 1$\,ps), the amplitude of the measured coefficients for large interaction energies are on the order of $|M_{21}| \sim 0.1 \sigma^2/\tau \sim 10^{-8}$\,m$^2$/s. 
This is comparable to the values reported in the recent experimental work of Bregulla et al. \cite{Bregulla2016}, on the order of $10^{-10}$ to $10^{-9}$\,m$^2$/s. 
Figure~\ref{fig:All_M_e} then reveals two interesting features. 
Firstly, the sign of $M_{21}$ changes around $\eps_{ls} = 0.8 \eps$: at low $\eps_{ls}$, the heat flux along the flow is positive, and at high $\eps_{ls}$ it is negative. 
Secondly, 
$M_{21}$ is strongly enhanced for the lowest liquid-solid interaction energies, reaching values up to $\sim 35 \sigma^2/\tau \sim 4 \times 10^{-6}$\,m$^2$/s.

\begin{figure}
	\centering
	\includegraphics[width=0.5\textwidth]{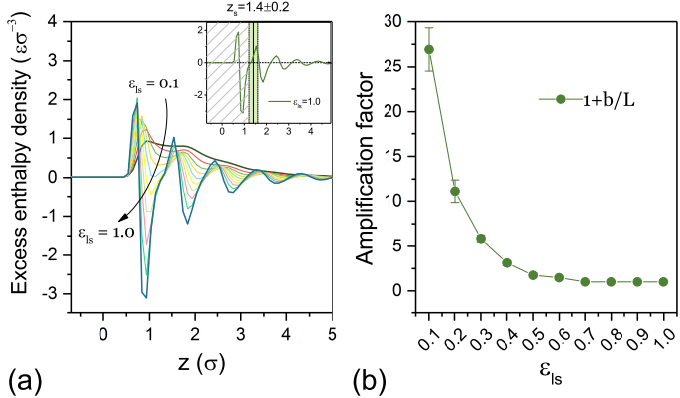}
	\caption{(a) Excess specific enthalpy profiles for different solid-liquid interaction energies $\eps_{ls}$; Inset: excess enthalpy density profile for $\eps_{ls}=1.0$; $z_s$ is the lower limit of the integration in Eq.~\ref{eq:M21_linear_slip} and has a large influence on the amplitude and sign of the thermo-osmosis coefficient. (b) Amplification factors for different solid-liquid interaction energies $\eps_{ls}$. All these data are measured using the mechano-caloric configuration.}
	\label{fig:enthalpy_amp}
\end{figure}

In order to understand the change of sign of the thermo-osmosis coefficient, Fig.~\ref{fig:enthalpy_amp}a shows the excess enthalpy profiles for different solid-liquid interaction energies. We note that for all the profiles, the excess enthalpy vanishes at $\sim 5 \sigma$ from the interface.
At the vicinity of the surface, the excess enthalpy is positive everywhere for $\eps_{ls}<0.6\eps$, which explains the positive value of $M_{21}$. For stronger interfacial interactions, the excess enthalpy \new{(corresponding to the driving force for thermo-osmosis)} shows large oscillations around zero, with a period on the order of the molecular size. \new{This is in strong contrast with electro- and diffusio-osmotic flows where the driving force generally does not change sign in the interfacial region}. According to Eq.~\ref{eq:M21_linear_slip}, $z_s$ fixes the lower limit of the integration, so that the value and also the sign of the response coefficient $M_{21}$ depend strongly on $z_s$, and cannot be predicted based only on equilibrium properties of the interface.
Overall these results shed some light on the structural and dynamical molecular mechanisms underlying the response coefficients of different signs that have been reported in the literature \cite{Derjaguin1980, Rusconi2004, Lusebrink2012, Nedev2015, Bregulla2016}.

However, the excess enthalpy profiles can not fully explain the massive enhancement of $M_{21}$ at low $\eps_{ls}$, and hydrodynamic slip must be taken into account. Following a previous treatment of slippage in diffusio-osmosis \cite{Ajdari:2006ci}, 
we introduce a characteristic length $L$ representative of the \new{thickness of the interfacial liquid layer} where the enthalpy differs from the bulk, 
\begin{equation}
L= \frac{\int_{z_s}^{+\infty} (z-z_s)\, \delta h(z)\df z}{\int_{z_s}^{+\infty}\delta h(z)\df z}.
\end{equation}
One can then rewrite Eq. \eqref{eq:M21_linear_slip}: 
\begin{equation}
M_{21} = M_{21}^{\mathrm{no-slip}} (1+b/L),
\label{eq:M21_Gamma_L_b} 
\end{equation}
where $M_{21}^{\mathrm{no-slip}}$ is the response that would be obtained in the absence of slip ($b=0$), and 
the amplification factor $1+b/L$ quantifies the contribution of hydrodynamic slip to the thermo-osmotic response. We plot in Fig.~\ref{fig:enthalpy_amp}b the evolution of the amplification factor against $\eps_{ls}$. For $\eps_{ls} > 0.6\eps$, the amplification factor converges to 1 since the slip length vanishes.
On the other hand, the amplification factor grows rapidly when $\eps_{ls}$ gets lower than $0.5\eps$, which shows that on hydrophobic surfaces the amplitude of the thermo-osmotic response is mostly controlled by slippage.  

\new{Such an amplification by hydrodynamic slip has already been reported for other osmotic flows \cite{Muller:1986wk, Joly2004, Joly2006, Bouzigues:2008db, Audry:2010gy, Ajdari:2006ci}. In particular, for electro-osmosis and diffusio-osmosis with electrolytes as solute, the amplification factor is controlled by the ratio between the slip length and the Debye length, quantifying the thickness of the electrical double layer. In moderately concentrated aqueous electrolytes, the Debye length can reach tens of nanometers, while here the thickness $L$ of the interfacial liquid layer is on the order of a few molecular sizes. Accordingly, the effect of slip is particularly large for thermo-osmosis.} 

\new{One should note finally that the two lowest $\eps_{ls}$ correspond to contact angles that cannot be achieved experimentally on smooth surfaces. Nevertheless, these $\eps_{ls}$ could be considered as an effective description of superhydrophobic (SH) surfaces, which display both very large contact angles and very high slip lengths \cite{Ma2006,Ybert2007}. The results obtained here therefore motivate further work on more realistic SH surfaces, which have been considered theoretically at the continuum level \cite{Baier2010}, but where molecular effects remain to be explored. We also emphasize that a significant amplification by slip is already observed for intermediate $\eps_{ls}$ values corresponding to more realistic smooth hydrophobic surfaces. Finally, identifying the  amplifying role of slip motivated us to consider a more realistic system exhibiting giant slippage, namely water/graphene, as detailed later.}

\begin{figure}
	\centering
	\includegraphics[width=0.5\textwidth]{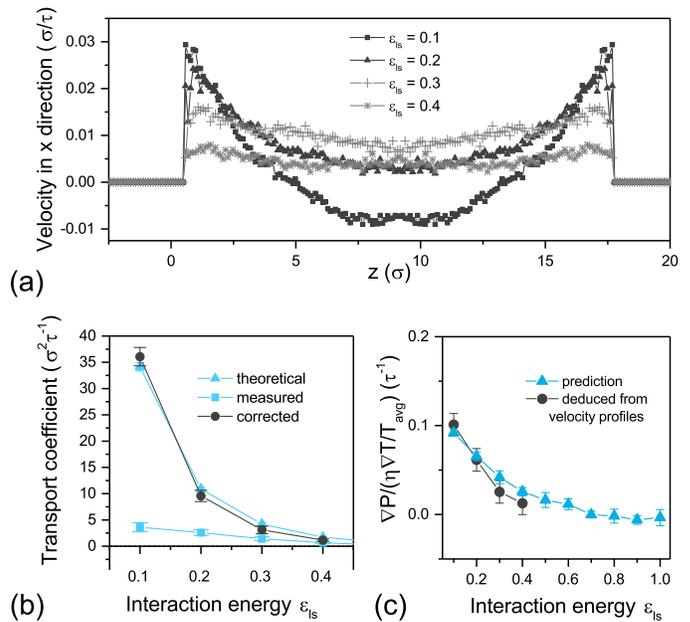}
	\caption{(a) Measured velocity profiles for the cases where a large disparity is observed between the theoretical and measured $M_{12}$ coefficient through the thermo-osmotic route. The lower the $\eps_{ls}$, the larger the curvature of the parabolic backflow. (b) Comparison between 3 estimates of the thermo-osmosis coefficient $M_{12}$: theoretical prediction, value deduced from raw velocity measurement, and value deduced from velocity corrected with the backflow velocity (see text for details). 
    (c) Evolution of the normalized pressure gradient responsible for the backflow as a function of the interaction energy. Blue triangle: derived from the correlation of theoretical and measured $M_{12}$; black circle: calculated from the curvature of velocity profiles according to Poiseuille law.}
	\label{fig:backflow}
\end{figure}

We now turn to the measurements of $M_{12}$ in the thermo-osmotic configuration,  
represented with blue closed symbols in Fig.~\ref{fig:All_M_e}. The theoretical $M_{12}$ matches the numerical and theoretical $M_{21}$ as expected from Onsager reciprocal relations. 
Also, both the theoretical and measured $M_{12}$ display a change of sign, corresponding to a reversal of the flow for a given thermal gradient direction. 
However, in contrast to what we found with the mechano-caloric route, there is a large discrepancy between the measured and theoretical $M_{12}$ at low $\eps_{ls}$: the massive increase predicted by the theory is systematically attenuated. In order to understand this
phenomenon, we plotted in Fig.~\ref{fig:backflow}a typical velocity profiles for low $\eps_{ls}$. Indeed, their parabolic shape reveals a Poiseuille backflow. We suggest this backflow is due to viscous entrance effects~\cite{Sampson:1891hj, Weissberg:1962iz, Dagan:1982wu}, whose key influence on nanoscale flows has been emphasized recently~\cite{Sisan2011, Joly2011, Gravelle2013, Walther2013, Gravelle:2014jj}. 
According to Poiseuille law, one can deduce the backflow velocity profile $v_{bk} (z)$ from the measured curvature of the velocity profile, using the hydrodynamic boundary condition  parameters determined in the mechano-caloric configuration. We then obtain the true thermo-osmotic velocity profile $v_s (z)$ by correcting the measured velocity $v_m (z)$ with the backflow velocity $v_{bk} (z)$: 
$v_{s}=v_m - v_{bk}$. Thus a corrected thermo-osmosis coefficient can be calculated from the definition, Eq.~\eqref{eq:coefficient_definition}, and the results are plotted in Fig.~\ref{fig:backflow}b. We found that the corrected values match well the theoretical ones. 
Regarding the cases for which the backflow is hardly detectable on the velocity profile, the pressure gradient responsible for the backflow can be estimated from the correlation of measured and theoretical thermo-osmosis coefficients. For a given system, we computed a normalized pressure gradient $\nabla P / (\eta \nabla T / T_{avg})$, which depends only on $\eps_{ls}$ (see the SM). The tendency is shown in Fig.~\ref{fig:backflow}c with blue closed triangles. The lower is $\eps_{ls}$, the higher is the pressure gradient, in line with what we obtained from the velocity profiles. We also deduced the same term from the curvature of the velocity profiles for those with evident backflow (typically $\eps_{ls} < 0.5$), as shown with the black circles in Fig.~\ref{fig:backflow}c. Within uncertainties, the thermo-osmosis coefficient correlation successfully predicts the pressure gradient generating the backflow. 

\paragraph*{Water-graphene interface} 
Finally, we estimated the thermo-osmosis coefficient of a water-graphene system using the mechano-caloric route, with a pressure of 1\,atm and a temperature of 323\,K (see the SM for details). We used the TIP4P/2005 force field for water~\cite{Abascal:2005ka}, the LCBOP one for graphene~\cite{Los:2003hd} and a recently proposed force field for water-carbon interactions~\cite{PerezHernandez:2013dr}, which has been shown to reproduce accurately quantum chemistry calculations of interaction energies between water and carbon nanostructures \cite{Al-Hamdani2017}. \new{With these force fields, we measured a contact angle of ca. $85\,^\circ$ and a slip length of ca. $32$\,nm, within the range of values reported in the literature \cite{Wang2016,Kannam:2013gf} (see details in the SM).} 
We obtained $M_{21} = (2.5\pm0.3) \times 10^{-6}$\,m$^2$/s. This value substantially exceeds those documented in other studies \cite{Bregulla2016, Derjaguin:1987bf} on the order of $10^{-10}$ to $10^{-9}$\,m$^2$/s. It is well known that the water-graphene interface presents a very low friction, and accordingly a very large slip length \cite{Kannam:2013gf}, largely exceeding the spatial range over which the specific enthalpy differs from its bulk value. 
In that limit, and using the relation between the slip length $b$ and the interfacial friction coefficient $\lambda$: $b = \eta/\lambda$ \cite{Bocquet2007}, Eq.~\eqref{eq:M21_linear_slip} can be simplified as $M_{21} \approx \frac{1}{\lambda} \int \delta h(z)\df z$, where the response coefficient only depends on the total enthalpy excess and on the interfacial friction coefficient. Indeed we were able to reproduce the measured $M_{21}$ using this formula. 
Note that the excess enthalpy profile of this system (see the SM) oscillates around zero and resembles a surface with intermediate wettability. Therefore, it is the very low friction coefficient which explains primarily the giant thermo-osmotic response. With an enhancement of about 3 orders of magnitude compared to existing experimental data, this water-graphene system shows a great potential for thermal energy harvesting\new{, although the very large thermal conductivity of graphene \cite{Balandin2008} could reduce the efficiency of such systems, an effect which we plan to investigate in the future}. 

\paragraph*{Summary}
\label{Summary}
We measured the thermo-osmosis coefficient using MD simulations via mechano-caloric and thermo-osmotic routes for different solid-liquid interfacial energies. 
A good agreement is obtained using these two methods, in line with Onsager reciprocal relations.
We showed that \new{the standard picture of thermo-osmosis as a thermal gradient-induced Marangoni flow can only give a qualitative description of the phenomenon, because it does not account for the pivotal role of} the hydrodynamic boundary condition. 
For high interfacial energies, due to the oscillations of the excess enthalpy profiles in the vicinity of the solid-liquid interface, the thermo-osmosis coefficient depends closely on the thickness of the stagnant liquid layer, and a change of sign is clearly observed. 
For low interfacial energies, hydrodynamic slip 
largely amplifies the thermo-osmosis coefficient, which has also been confirmed by simulations of the water-graphene interface. 
Finally we showed that viscous entrance effects reduce significantly the amplitude of the thermo-osmotic flow in the considered nanopore geometries. 
We hope these results will motivate future experimental development and characterization of new functional interfaces for efficient harvesting of thermal energy.

\begin{acknowledgments}
The authors thank A.-L. Biance, O. Bonhomme, C. Ybert and C. Cottin-Bizonne for fruitful discussions. This work is supported by the ANR, project ANR-16-CE06-0004-01 NECtAR. 
\end{acknowledgments}

\bibliography{bibliography,Mendeley}

\begin{thebibliography}{80}%
\makeatletter
\providecommand \@ifxundefined [1]{%
 \@ifx{#1\undefined}
}%
\providecommand \@ifnum [1]{%
 \ifnum #1\expandafter \@firstoftwo
 \else \expandafter \@secondoftwo
 \fi
}%
\providecommand \@ifx [1]{%
 \ifx #1\expandafter \@firstoftwo
 \else \expandafter \@secondoftwo
 \fi
}%
\providecommand \natexlab [1]{#1}%
\providecommand \enquote  [1]{``#1''}%
\providecommand \bibnamefont  [1]{#1}%
\providecommand \bibfnamefont [1]{#1}%
\providecommand \citenamefont [1]{#1}%
\providecommand \href@noop [0]{\@secondoftwo}%
\providecommand \href [0]{\begingroup \@sanitize@url \@href}%
\providecommand \@href[1]{\@@startlink{#1}\@@href}%
\providecommand \@@href[1]{\endgroup#1\@@endlink}%
\providecommand \@sanitize@url [0]{\catcode `\\12\catcode `\$12\catcode
  `\&12\catcode `\#12\catcode `\^12\catcode `\_12\catcode `\%12\relax}%
\providecommand \@@startlink[1]{}%
\providecommand \@@endlink[0]{}%
\providecommand \url  [0]{\begingroup\@sanitize@url \@url }%
\providecommand \@url [1]{\endgroup\@href {#1}{\urlprefix }}%
\providecommand \urlprefix  [0]{URL }%
\providecommand \Eprint [0]{\href }%
\providecommand \doibase [0]{http://dx.doi.org/}%
\providecommand \selectlanguage [0]{\@gobble}%
\providecommand \bibinfo  [0]{\@secondoftwo}%
\providecommand \bibfield  [0]{\@secondoftwo}%
\providecommand \translation [1]{[#1]}%
\providecommand \BibitemOpen [0]{}%
\providecommand \bibitemStop [0]{}%
\providecommand \bibitemNoStop [0]{.\EOS\space}%
\providecommand \EOS [0]{\spacefactor3000\relax}%
\providecommand \BibitemShut  [1]{\csname bibitem#1\endcsname}%
\let\auto@bib@innerbib\@empty
\bibitem [{\citenamefont {Golestanian}\ \emph {et~al.}(2007)\citenamefont
  {Golestanian}, \citenamefont {Liverpool},\ and\ \citenamefont
  {Ajdari}}]{Golestanian2007}%
  \BibitemOpen
  \bibfield  {author} {\bibinfo {author} {\bibfnamefont {R.}~\bibnamefont
  {Golestanian}}, \bibinfo {author} {\bibfnamefont {T.~B.}\ \bibnamefont
  {Liverpool}}, \ and\ \bibinfo {author} {\bibfnamefont {A.}~\bibnamefont
  {Ajdari}},\ }\href {\doibase 10.1088/1367-2630/9/5/126} {\bibfield  {journal}
  {\bibinfo  {journal} {New Journal of Physics}\ }\textbf {\bibinfo {volume}
  {9}},\ \bibinfo {pages} {126} (\bibinfo {year} {2007})}\BibitemShut {NoStop}%
\bibitem [{\citenamefont {Dhont}\ \emph {et~al.}(2007)\citenamefont {Dhont},
  \citenamefont {Wiegand}, \citenamefont {Duhr},\ and\ \citenamefont
  {Braun}}]{Dhont2007}%
  \BibitemOpen
  \bibfield  {author} {\bibinfo {author} {\bibfnamefont {J.~K.~G.}\
  \bibnamefont {Dhont}}, \bibinfo {author} {\bibfnamefont {S.}~\bibnamefont
  {Wiegand}}, \bibinfo {author} {\bibfnamefont {S.}~\bibnamefont {Duhr}}, \
  and\ \bibinfo {author} {\bibfnamefont {D.}~\bibnamefont {Braun}},\ }\href
  {\doibase 10.1021/la062184m} {\bibfield  {journal} {\bibinfo  {journal}
  {Langmuir}\ }\textbf {\bibinfo {volume} {23}},\ \bibinfo {pages} {1674}
  (\bibinfo {year} {2007})}\BibitemShut {NoStop}%
\bibitem [{\citenamefont {Piazza}\ and\ \citenamefont
  {Parola}(2008)}]{Piazza2008}%
  \BibitemOpen
  \bibfield  {author} {\bibinfo {author} {\bibfnamefont {R.}~\bibnamefont
  {Piazza}}\ and\ \bibinfo {author} {\bibfnamefont {A.}~\bibnamefont
  {Parola}},\ }\href {\doibase 10.1088/0953-8984/20/15/153102} {\bibfield
  {journal} {\bibinfo  {journal} {Journal of Physics: Condensed Matter}\
  }\textbf {\bibinfo {volume} {20}},\ \bibinfo {pages} {153102} (\bibinfo
  {year} {2008})}\BibitemShut {NoStop}%
\bibitem [{\citenamefont {Zambrano}\ \emph {et~al.}(2009)\citenamefont
  {Zambrano}, \citenamefont {Walther}, \citenamefont {Koumoutsakos},\ and\
  \citenamefont {Sbalzarini}}]{Zambrano2009}%
  \BibitemOpen
  \bibfield  {author} {\bibinfo {author} {\bibfnamefont {H.~A.}\ \bibnamefont
  {Zambrano}}, \bibinfo {author} {\bibfnamefont {J.~H.}\ \bibnamefont
  {Walther}}, \bibinfo {author} {\bibfnamefont {P.}~\bibnamefont
  {Koumoutsakos}}, \ and\ \bibinfo {author} {\bibfnamefont {I.~F.}\
  \bibnamefont {Sbalzarini}},\ }\href {\doibase 10.1021/nl802429s} {\bibfield
  {journal} {\bibinfo  {journal} {Nano Letters}\ }\textbf {\bibinfo {volume}
  {9}},\ \bibinfo {pages} {66} (\bibinfo {year} {2009})}\BibitemShut {NoStop}%
\bibitem [{\citenamefont {Shiomi}\ and\ \citenamefont
  {Maruyama}(2009)}]{Shiomi2009}%
  \BibitemOpen
  \bibfield  {author} {\bibinfo {author} {\bibfnamefont {J.}~\bibnamefont
  {Shiomi}}\ and\ \bibinfo {author} {\bibfnamefont {S.}~\bibnamefont
  {Maruyama}},\ }\href {\doibase 10.1088/0957-4484/20/5/055708} {\bibfield
  {journal} {\bibinfo  {journal} {Nanotechnology}\ }\textbf {\bibinfo {volume}
  {20}},\ \bibinfo {pages} {055708} (\bibinfo {year} {2009})}\BibitemShut
  {NoStop}%
\bibitem [{\citenamefont {W\"urger}(2010)}]{Wurger:2010dj}%
  \BibitemOpen
  \bibfield  {author} {\bibinfo {author} {\bibfnamefont {A.}~\bibnamefont
  {W\"urger}},\ }\href {\doibase 10.1088/0034-4885/73/12/126601} {\bibfield
  {journal} {\bibinfo  {journal} {Reports on Progress in Physics}\ }\textbf
  {\bibinfo {volume} {73}},\ \bibinfo {pages} {126601} (\bibinfo {year}
  {2010})}\BibitemShut {NoStop}%
\bibitem [{\citenamefont {Wolff}\ \emph {et~al.}(2016)\citenamefont {Wolff},
  \citenamefont {Mittag}, \citenamefont {Herling}, \citenamefont {De~Genst},
  \citenamefont {Dobson}, \citenamefont {Knowles}, \citenamefont {Braun},\ and\
  \citenamefont {Buell}}]{Wolff:2016fh}%
  \BibitemOpen
  \bibfield  {author} {\bibinfo {author} {\bibfnamefont {M.}~\bibnamefont
  {Wolff}}, \bibinfo {author} {\bibfnamefont {J.~J.}\ \bibnamefont {Mittag}},
  \bibinfo {author} {\bibfnamefont {T.~W.}\ \bibnamefont {Herling}}, \bibinfo
  {author} {\bibfnamefont {E.}~\bibnamefont {De~Genst}}, \bibinfo {author}
  {\bibfnamefont {C.~M.}\ \bibnamefont {Dobson}}, \bibinfo {author}
  {\bibfnamefont {T.~P.~J.}\ \bibnamefont {Knowles}}, \bibinfo {author}
  {\bibfnamefont {D.}~\bibnamefont {Braun}}, \ and\ \bibinfo {author}
  {\bibfnamefont {A.~K.}\ \bibnamefont {Buell}},\ }\href {\doibase
  10.1038/srep22829} {\bibfield  {journal} {\bibinfo  {journal} {Scientific
  reports}\ }\textbf {\bibinfo {volume} {6}},\ \bibinfo {pages} {22829}
  (\bibinfo {year} {2016})}\BibitemShut {NoStop}%
\bibitem [{\citenamefont {Rajegowda}\ \emph {et~al.}(2017)\citenamefont
  {Rajegowda}, \citenamefont {Kannam}, \citenamefont {Hartkamp},\ and\
  \citenamefont {Sathian}}]{Rajegowda2017}%
  \BibitemOpen
  \bibfield  {author} {\bibinfo {author} {\bibfnamefont {R.}~\bibnamefont
  {Rajegowda}}, \bibinfo {author} {\bibfnamefont {S.~K.}\ \bibnamefont
  {Kannam}}, \bibinfo {author} {\bibfnamefont {R.}~\bibnamefont {Hartkamp}}, \
  and\ \bibinfo {author} {\bibfnamefont {S.~P.}\ \bibnamefont {Sathian}},\
  }\href {\doibase 10.1088/1361-6528/aa6290} {\bibfield  {journal} {\bibinfo
  {journal} {Nanotechnology}\ }\textbf {\bibinfo {volume} {28}},\ \bibinfo
  {pages} {155401} (\bibinfo {year} {2017})}\BibitemShut {NoStop}%
\bibitem [{\citenamefont {Demirel}\ and\ \citenamefont
  {Sandler}(2001)}]{Demirel2001}%
  \BibitemOpen
  \bibfield  {author} {\bibinfo {author} {\bibfnamefont {Y.}~\bibnamefont
  {Demirel}}\ and\ \bibinfo {author} {\bibfnamefont {S.~I.}\ \bibnamefont
  {Sandler}},\ }\href@noop {} {\bibfield  {journal} {\bibinfo  {journal}
  {International Journal of Heat and Mass Transfer}\ }\textbf {\bibinfo
  {volume} {44}},\ \bibinfo {pages} {2439} (\bibinfo {year}
  {2001})}\BibitemShut {NoStop}%
\bibitem [{\citenamefont {Parola}\ and\ \citenamefont
  {Piazza}(2004)}]{Parola2004}%
  \BibitemOpen
  \bibfield  {author} {\bibinfo {author} {\bibfnamefont {A.}~\bibnamefont
  {Parola}}\ and\ \bibinfo {author} {\bibfnamefont {R.}~\bibnamefont
  {Piazza}},\ }\href@noop {} {\bibfield  {journal} {\bibinfo  {journal} {The
  European physical journal. E, Soft matter}\ }\textbf {\bibinfo {volume}
  {15}},\ \bibinfo {pages} {255} (\bibinfo {year} {2004})}\BibitemShut
  {NoStop}%
\bibitem [{\citenamefont {W\"urger}(2007)}]{Wurger2007}%
  \BibitemOpen
  \bibfield  {author} {\bibinfo {author} {\bibfnamefont {A.}~\bibnamefont
  {W\"urger}},\ }\href@noop {} {\bibfield  {journal} {\bibinfo  {journal}
  {Physical Review Letters}\ }\textbf {\bibinfo {volume} {98}},\ \bibinfo
  {pages} {138301} (\bibinfo {year} {2007})}\BibitemShut {NoStop}%
\bibitem [{\citenamefont {Dhont}\ and\ \citenamefont
  {Briels}(2008)}]{Dhont2008}%
  \BibitemOpen
  \bibfield  {author} {\bibinfo {author} {\bibfnamefont {J.~K.~G.}\
  \bibnamefont {Dhont}}\ and\ \bibinfo {author} {\bibfnamefont {W.~J.}\
  \bibnamefont {Briels}},\ }\href@noop {} {\bibfield  {journal} {\bibinfo
  {journal} {The European physical journal. E, Soft matter}\ }\textbf {\bibinfo
  {volume} {25}},\ \bibinfo {pages} {61} (\bibinfo {year} {2008})}\BibitemShut
  {NoStop}%
\bibitem [{\citenamefont {Burelbach}\ \emph {et~al.}()\citenamefont
  {Burelbach}, \citenamefont {Frenkel}, \citenamefont {Pagonabarraga},\ and\
  \citenamefont {Eiser}}]{Burelbach2017}%
  \BibitemOpen
  \bibfield  {author} {\bibinfo {author} {\bibfnamefont {J.}~\bibnamefont
  {Burelbach}}, \bibinfo {author} {\bibfnamefont {D.}~\bibnamefont {Frenkel}},
  \bibinfo {author} {\bibfnamefont {I.}~\bibnamefont {Pagonabarraga}}, \ and\
  \bibinfo {author} {\bibfnamefont {E.}~\bibnamefont {Eiser}},\ }\href
  {http://arxiv.org/abs/1708.02195} {\bibinfo  {journal} {arXiv:1708.02195
  [cond-mat.soft]}\ }\BibitemShut {NoStop}%
\bibitem [{\citenamefont {Joly}\ \emph {et~al.}(2004)\citenamefont {Joly},
  \citenamefont {Ybert}, \citenamefont {Trizac},\ and\ \citenamefont
  {Bocquet}}]{Joly2004}%
  \BibitemOpen
\bibfield  {journal} {  }\bibfield  {author} {\bibinfo {author} {\bibfnamefont
  {L.}~\bibnamefont {Joly}}, \bibinfo {author} {\bibfnamefont {C.}~\bibnamefont
  {Ybert}}, \bibinfo {author} {\bibfnamefont {E.}~\bibnamefont {Trizac}}, \
  and\ \bibinfo {author} {\bibfnamefont {L.}~\bibnamefont {Bocquet}},\ }\href
  {\doibase 10.1103/PhysRevLett.93.257805} {\bibfield  {journal} {\bibinfo
  {journal} {Phys. Rev. Lett.}\ }\textbf {\bibinfo {volume} {93}},\ \bibinfo
  {pages} {257805} (\bibinfo {year} {2004})}\BibitemShut {NoStop}%
\bibitem [{\citenamefont {Huang}\ \emph {et~al.}(2007)\citenamefont {Huang},
  \citenamefont {Cottin-Bizonne}, \citenamefont {Ybert},\ and\ \citenamefont
  {Bocquet}}]{Huang2007}%
  \BibitemOpen
  \bibfield  {author} {\bibinfo {author} {\bibfnamefont {D.}~\bibnamefont
  {Huang}}, \bibinfo {author} {\bibfnamefont {C.}~\bibnamefont
  {Cottin-Bizonne}}, \bibinfo {author} {\bibfnamefont {C.}~\bibnamefont
  {Ybert}}, \ and\ \bibinfo {author} {\bibfnamefont {L.}~\bibnamefont
  {Bocquet}},\ }\href {\doibase 10.1103/PhysRevLett.98.177801} {\bibfield
  {journal} {\bibinfo  {journal} {Physical Review Letters}\ }\textbf {\bibinfo
  {volume} {98}},\ \bibinfo {pages} {177801} (\bibinfo {year}
  {2007})}\BibitemShut {NoStop}%
\bibitem [{\citenamefont {Rotenberg}\ and\ \citenamefont
  {Pagonabarraga}(2013)}]{Rotenberg2013}%
  \BibitemOpen
  \bibfield  {author} {\bibinfo {author} {\bibfnamefont {B.}~\bibnamefont
  {Rotenberg}}\ and\ \bibinfo {author} {\bibfnamefont {I.}~\bibnamefont
  {Pagonabarraga}},\ }\href {\doibase 10.1080/00268976.2013.791731} {\bibfield
  {journal} {\bibinfo  {journal} {Molecular Physics}\ }\textbf {\bibinfo
  {volume} {111}},\ \bibinfo {pages} {827} (\bibinfo {year}
  {2013})}\BibitemShut {NoStop}%
\bibitem [{\citenamefont {Bonthuis}\ and\ \citenamefont
  {Netz}(2013)}]{Bonthuis2013}%
  \BibitemOpen
  \bibfield  {author} {\bibinfo {author} {\bibfnamefont {D.~J.}\ \bibnamefont
  {Bonthuis}}\ and\ \bibinfo {author} {\bibfnamefont {R.~R.}\ \bibnamefont
  {Netz}},\ }\href {\doibase 10.1021/jp402482q} {\bibfield  {journal} {\bibinfo
   {journal} {The Journal of Physical Chemistry B}\ }\textbf {\bibinfo {volume}
  {117}},\ \bibinfo {pages} {11397} (\bibinfo {year} {2013})}\BibitemShut
  {NoStop}%
\bibitem [{\citenamefont {Majumder}\ \emph {et~al.}(2015)\citenamefont
  {Majumder}, \citenamefont {Dhar},\ and\ \citenamefont
  {Chakraborty}}]{Majumder2015}%
  \BibitemOpen
  \bibfield  {author} {\bibinfo {author} {\bibfnamefont {S.}~\bibnamefont
  {Majumder}}, \bibinfo {author} {\bibfnamefont {J.}~\bibnamefont {Dhar}}, \
  and\ \bibinfo {author} {\bibfnamefont {S.}~\bibnamefont {Chakraborty}},\
  }\href {\doibase 10.1038/srep14725} {\bibfield  {journal} {\bibinfo
  {journal} {Sci. Rep.}\ }\textbf {\bibinfo {volume} {5}},\ \bibinfo {pages}
  {14725} (\bibinfo {year} {2015})}\BibitemShut {NoStop}%
\bibitem [{\citenamefont {Maduar}\ \emph {et~al.}(2015)\citenamefont {Maduar},
  \citenamefont {Belyaev}, \citenamefont {Lobaskin},\ and\ \citenamefont
  {Vinogradova}}]{Maduar2015}%
  \BibitemOpen
  \bibfield  {author} {\bibinfo {author} {\bibfnamefont {S.~R.}\ \bibnamefont
  {Maduar}}, \bibinfo {author} {\bibfnamefont {A.~V.}\ \bibnamefont {Belyaev}},
  \bibinfo {author} {\bibfnamefont {V.}~\bibnamefont {Lobaskin}}, \ and\
  \bibinfo {author} {\bibfnamefont {O.~I.}\ \bibnamefont {Vinogradova}},\
  }\href {\doibase 10.1103/PhysRevLett.114.118301} {\bibfield  {journal}
  {\bibinfo  {journal} {Physical Review Letters}\ }\textbf {\bibinfo {volume}
  {114}},\ \bibinfo {pages} {118301} (\bibinfo {year} {2015})}\BibitemShut
  {NoStop}%
\bibitem [{\citenamefont {P{\v{r}}edota}\ \emph {et~al.}(2016)\citenamefont
  {P{\v{r}}edota}, \citenamefont {Machesky},\ and\ \citenamefont
  {Wesolowski}}]{Predota2016}%
  \BibitemOpen
  \bibfield  {author} {\bibinfo {author} {\bibfnamefont {M.}~\bibnamefont
  {P{\v{r}}edota}}, \bibinfo {author} {\bibfnamefont {M.~L.}\ \bibnamefont
  {Machesky}}, \ and\ \bibinfo {author} {\bibfnamefont {D.~J.}\ \bibnamefont
  {Wesolowski}},\ }\href {\doibase 10.1021/acs.langmuir.6b02493} {\bibfield
  {journal} {\bibinfo  {journal} {Langmuir}\ }\textbf {\bibinfo {volume}
  {32}},\ \bibinfo {pages} {10189} (\bibinfo {year} {2016})}\BibitemShut
  {NoStop}%
\bibitem [{\citenamefont {Siboulet}\ \emph {et~al.}(2017)\citenamefont
  {Siboulet}, \citenamefont {Hocine}, \citenamefont {Hartkamp},\ and\
  \citenamefont {Dufr{\^{e}}che}}]{Siboulet2017}%
  \BibitemOpen
  \bibfield  {author} {\bibinfo {author} {\bibfnamefont {B.}~\bibnamefont
  {Siboulet}}, \bibinfo {author} {\bibfnamefont {S.}~\bibnamefont {Hocine}},
  \bibinfo {author} {\bibfnamefont {R.}~\bibnamefont {Hartkamp}}, \ and\
  \bibinfo {author} {\bibfnamefont {J.-F.}\ \bibnamefont {Dufr{\^{e}}che}},\
  }\href {\doibase 10.1021/acs.jpcc.7b00309} {\bibfield  {journal} {\bibinfo
  {journal} {The Journal of Physical Chemistry C}\ }\textbf {\bibinfo {volume}
  {121}},\ \bibinfo {pages} {6756} (\bibinfo {year} {2017})}\BibitemShut
  {NoStop}%
\bibitem [{\citenamefont {Bhadauria}\ and\ \citenamefont
  {Aluru}(2017)}]{Bhadauria2017}%
  \BibitemOpen
  \bibfield  {author} {\bibinfo {author} {\bibfnamefont {R.}~\bibnamefont
  {Bhadauria}}\ and\ \bibinfo {author} {\bibfnamefont {N.~R.}\ \bibnamefont
  {Aluru}},\ }\href {\doibase 10.1063/1.4982731} {\bibfield  {journal}
  {\bibinfo  {journal} {The Journal of Chemical Physics}\ }\textbf {\bibinfo
  {volume} {146}},\ \bibinfo {pages} {184106} (\bibinfo {year}
  {2017})}\BibitemShut {NoStop}%
\bibitem [{\citenamefont {Ajdari}\ and\ \citenamefont
  {Bocquet}(2006)}]{Ajdari:2006ci}%
  \BibitemOpen
  \bibfield  {author} {\bibinfo {author} {\bibfnamefont {A.}~\bibnamefont
  {Ajdari}}\ and\ \bibinfo {author} {\bibfnamefont {L.}~\bibnamefont
  {Bocquet}},\ }\href {\doibase 10.1103/PhysRevLett.96.186102} {\bibfield
  {journal} {\bibinfo  {journal} {Physical Review Letters}\ }\textbf {\bibinfo
  {volume} {96}},\ \bibinfo {pages} {186102} (\bibinfo {year}
  {2006})}\BibitemShut {NoStop}%
\bibitem [{\citenamefont {Marbach}\ \emph {et~al.}(2017)\citenamefont
  {Marbach}, \citenamefont {Yoshida},\ and\ \citenamefont
  {Bocquet}}]{Marbach2017}%
  \BibitemOpen
  \bibfield  {author} {\bibinfo {author} {\bibfnamefont {S.}~\bibnamefont
  {Marbach}}, \bibinfo {author} {\bibfnamefont {H.}~\bibnamefont {Yoshida}}, \
  and\ \bibinfo {author} {\bibfnamefont {L.}~\bibnamefont {Bocquet}},\ }\href
  {\doibase 10.1063/1.4982221} {\bibfield  {journal} {\bibinfo  {journal} {The
  Journal of Chemical Physics}\ }\textbf {\bibinfo {volume} {146}},\ \bibinfo
  {pages} {194701} (\bibinfo {year} {2017})}\BibitemShut {NoStop}%
\bibitem [{\citenamefont {Yoshida}\ \emph {et~al.}(2017)\citenamefont
  {Yoshida}, \citenamefont {Marbach},\ and\ \citenamefont
  {Bocquet}}]{Yoshida2017}%
  \BibitemOpen
  \bibfield  {author} {\bibinfo {author} {\bibfnamefont {H.}~\bibnamefont
  {Yoshida}}, \bibinfo {author} {\bibfnamefont {S.}~\bibnamefont {Marbach}}, \
  and\ \bibinfo {author} {\bibfnamefont {L.}~\bibnamefont {Bocquet}},\ }\href
  {\doibase 10.1063/1.4981794} {\bibfield  {journal} {\bibinfo  {journal} {The
  Journal of Chemical Physics}\ }\textbf {\bibinfo {volume} {146}},\ \bibinfo
  {pages} {194702} (\bibinfo {year} {2017})}\BibitemShut {NoStop}%
\bibitem [{\citenamefont {Lee}\ \emph {et~al.}(2017)\citenamefont {Lee},
  \citenamefont {Cottin-Bizonne}, \citenamefont {Fulcrand}, \citenamefont
  {Joly},\ and\ \citenamefont {Ybert}}]{Lee:2017cw}%
  \BibitemOpen
  \bibfield  {author} {\bibinfo {author} {\bibfnamefont {C.}~\bibnamefont
  {Lee}}, \bibinfo {author} {\bibfnamefont {C.}~\bibnamefont {Cottin-Bizonne}},
  \bibinfo {author} {\bibfnamefont {R.}~\bibnamefont {Fulcrand}}, \bibinfo
  {author} {\bibfnamefont {L.}~\bibnamefont {Joly}}, \ and\ \bibinfo {author}
  {\bibfnamefont {C.}~\bibnamefont {Ybert}},\ }\href {\doibase
  10.1021/acs.jpclett.6b02753} {\bibfield  {journal} {\bibinfo  {journal} {The
  Journal of Physical Chemistry Letters}\ }\textbf {\bibinfo {volume} {8}},\
  \bibinfo {pages} {478–483} (\bibinfo {year} {2017})}\BibitemShut {NoStop}%
\bibitem [{\citenamefont {Derjaguin}\ and\ \citenamefont
  {Sidorenkov}(1941)}]{Derjaguin:1941vs}%
  \BibitemOpen
  \bibfield  {author} {\bibinfo {author} {\bibfnamefont {B.}~\bibnamefont
  {Derjaguin}}\ and\ \bibinfo {author} {\bibfnamefont {G.~P.}\ \bibnamefont
  {Sidorenkov}},\ }\href@noop {} {\emph {\bibinfo {title} {On thermo-osmosis of
  liquid in porous glass}}},\ Vol.~\bibinfo {volume} {32}\ (\bibinfo
  {publisher} {CR Acad. Sci. URSS},\ \bibinfo {year} {1941})\ p.\ \bibinfo
  {pages} {622–626}\BibitemShut {NoStop}%
\bibitem [{\citenamefont {Ruckenstein}(1981)}]{Ruckenstein:1981kg}%
  \BibitemOpen
  \bibfield  {author} {\bibinfo {author} {\bibfnamefont {E.}~\bibnamefont
  {Ruckenstein}},\ }\href@noop {} {\bibfield  {journal} {\bibinfo  {journal} {J
  Colloid Interface Sci}\ }\textbf {\bibinfo {volume} {83}},\ \bibinfo {pages}
  {77} (\bibinfo {year} {1981})}\BibitemShut {NoStop}%
\bibitem [{\citenamefont {Anderson}(1989)}]{Anderson:1989vq}%
  \BibitemOpen
  \bibfield  {author} {\bibinfo {author} {\bibfnamefont {J.~L.}\ \bibnamefont
  {Anderson}},\ }\href@noop {} {\bibfield  {journal} {\bibinfo  {journal}
  {Annual review of fluid mechanics}\ } (\bibinfo {year} {1989})}\BibitemShut
  {NoStop}%
\bibitem [{\citenamefont {Derjaguin}\ \emph {et~al.}(1987)\citenamefont
  {Derjaguin}, \citenamefont {Churaev},\ and\ \citenamefont
  {Muller}}]{Derjaguin:1987bf}%
  \BibitemOpen
  \bibfield  {author} {\bibinfo {author} {\bibfnamefont {B.~V.}\ \bibnamefont
  {Derjaguin}}, \bibinfo {author} {\bibfnamefont {N.~V.}\ \bibnamefont
  {Churaev}}, \ and\ \bibinfo {author} {\bibfnamefont {V.~M.}\ \bibnamefont
  {Muller}},\ }\enquote {\bibinfo {title} {Surface forces in transport
  phenomena},}\ in\ \href {\doibase 10.1007/978-1-4757-6639-4_11} {\emph
  {\bibinfo {booktitle} {Surface Forces}}}\ (\bibinfo  {publisher} {Springer
  US},\ \bibinfo {address} {Boston, MA},\ \bibinfo {year} {1987})\ pp.\
  \bibinfo {pages} {369--431}\BibitemShut {NoStop}%
\bibitem [{\citenamefont {Dietzel}\ and\ \citenamefont
  {Hardt}(2017)}]{Dietzel2017}%
  \BibitemOpen
  \bibfield  {author} {\bibinfo {author} {\bibfnamefont {M.}~\bibnamefont
  {Dietzel}}\ and\ \bibinfo {author} {\bibfnamefont {S.}~\bibnamefont
  {Hardt}},\ }\href {\doibase 10.1017/jfm.2016.844} {\bibfield  {journal}
  {\bibinfo  {journal} {Journal of Fluid Mechanics}\ }\textbf {\bibinfo
  {volume} {813}},\ \bibinfo {pages} {1060} (\bibinfo {year}
  {2017})}\BibitemShut {NoStop}%
\bibitem [{\citenamefont {Ganti}\ \emph {et~al.}(2017)\citenamefont {Ganti},
  \citenamefont {Liu},\ and\ \citenamefont {Frenkel}}]{Ganti2017}%
  \BibitemOpen
  \bibfield  {author} {\bibinfo {author} {\bibfnamefont {R.}~\bibnamefont
  {Ganti}}, \bibinfo {author} {\bibfnamefont {Y.}~\bibnamefont {Liu}}, \ and\
  \bibinfo {author} {\bibfnamefont {D.}~\bibnamefont {Frenkel}},\ }\href
  {\doibase 10.1103/PhysRevLett.119.038002} {\bibfield  {journal} {\bibinfo
  {journal} {Physical Review Letters}\ }\textbf {\bibinfo {volume} {119}},\
  \bibinfo {pages} {038002} (\bibinfo {year} {2017})}\BibitemShut {NoStop}%
\bibitem [{\citenamefont {Bregulla}\ \emph {et~al.}(2016)\citenamefont
  {Bregulla}, \citenamefont {W{\"{u}}rger}, \citenamefont {G{\"{u}}nther},
  \citenamefont {Mertig},\ and\ \citenamefont {Cichos}}]{Bregulla2016}%
  \BibitemOpen
  \bibfield  {author} {\bibinfo {author} {\bibfnamefont {A.~P.}\ \bibnamefont
  {Bregulla}}, \bibinfo {author} {\bibfnamefont {A.}~\bibnamefont
  {W{\"{u}}rger}}, \bibinfo {author} {\bibfnamefont {K.}~\bibnamefont
  {G{\"{u}}nther}}, \bibinfo {author} {\bibfnamefont {M.}~\bibnamefont
  {Mertig}}, \ and\ \bibinfo {author} {\bibfnamefont {F.}~\bibnamefont
  {Cichos}},\ }\href {\doibase 10.1103/PhysRevLett.116.188303} {\bibfield
  {journal} {\bibinfo  {journal} {Physical Review Letters}\ }\textbf {\bibinfo
  {volume} {116}},\ \bibinfo {pages} {188303} (\bibinfo {year}
  {2016})}\BibitemShut {NoStop}%
\bibitem [{\citenamefont {Kjelstrup}\ and\ \citenamefont
  {Bedeaux}(2008)}]{Kjelstrup:2008iu}%
  \BibitemOpen
  \bibfield  {author} {\bibinfo {author} {\bibfnamefont {S.}~\bibnamefont
  {Kjelstrup}}\ and\ \bibinfo {author} {\bibfnamefont {D.}~\bibnamefont
  {Bedeaux}},\ }\href {\doibase 10.1142/9789812779144} {\emph {\bibinfo {title}
  {Non-equilibrium thermodynamics of heterogeneous systems}}},\ Vol.~\bibinfo
  {volume} {16}\ (\bibinfo  {publisher} {World Scientific Publishing Co. Pte.
  Ltd., Hackensack, NJ},\ \bibinfo {year} {2008})\BibitemShut {NoStop}%
\bibitem [{\citenamefont {de~Groot}\ and\ \citenamefont
  {Mazur}(1984)}]{deGroot:1984ue}%
  \BibitemOpen
  \bibfield  {author} {\bibinfo {author} {\bibfnamefont {S.~R.}\ \bibnamefont
  {de~Groot}}\ and\ \bibinfo {author} {\bibfnamefont {P.}~\bibnamefont
  {Mazur}},\ }\href@noop {} {\emph {\bibinfo {title} {Non-equilibrium
  Thermodynamics}}}\ (\bibinfo  {publisher} {Dover Publications},\ \bibinfo
  {year} {1984})\BibitemShut {NoStop}%
\bibitem [{\citenamefont {Brunet}\ and\ \citenamefont
  {Ajdari}(2004)}]{Brunet:2004bv}%
  \BibitemOpen
  \bibfield  {author} {\bibinfo {author} {\bibfnamefont {E.}~\bibnamefont
  {Brunet}}\ and\ \bibinfo {author} {\bibfnamefont {A.}~\bibnamefont
  {Ajdari}},\ }\href {\doibase 10.1103/PhysRevE.69.016306} {\bibfield
  {journal} {\bibinfo  {journal} {Physical Review E}\ }\textbf {\bibinfo
  {volume} {69}},\ \bibinfo {pages} {016306} (\bibinfo {year}
  {2004})}\BibitemShut {NoStop}%
\bibitem [{Note1()}]{Note1}%
  \BibitemOpen
  \bibinfo {note} {See Supplemental Material at [url] for further details,
  which includes Refs. \cite
  {Jewett:2013fg,Humphrey:1996bk,Bussi2007,Barrat:1999kv,Evans:2016ur,Tocci2014a,Falk2010,KumarKannam2012,Guevara-Carrion2011,Guillaud:2017bk,Markesteijn2012}.}\BibitemShut
  {Stop}%
\bibitem [{\citenamefont {Han}(2005)}]{Han:2005ig}%
  \BibitemOpen
  \bibfield  {author} {\bibinfo {author} {\bibfnamefont {M.}~\bibnamefont
  {Han}},\ }\href {\doibase 10.1016/j.jcis.2004.09.067} {\bibfield  {journal}
  {\bibinfo  {journal} {J Colloid Interface Sci}\ }\textbf {\bibinfo {volume}
  {284}},\ \bibinfo {pages} {339–348} (\bibinfo {year} {2005})}\BibitemShut
  {NoStop}%
\bibitem [{\citenamefont {Muller}\ and\ \citenamefont
  {Sergeeva}(1986)}]{Muller:1986wk}%
  \BibitemOpen
  \bibfield  {author} {\bibinfo {author} {\bibfnamefont {V.~M.}\ \bibnamefont
  {Muller}}\ and\ \bibinfo {author} {\bibfnamefont {I.~P.}\ \bibnamefont
  {Sergeeva}},\ }\href@noop {} {\emph {\bibinfo {title} {Boundary effects in
  the theory of electrokinetic phenomena}}}\ (\bibinfo  {publisher} {…
  Journal of the …},\ \bibinfo {year} {1986})\BibitemShut {NoStop}%
\bibitem [{\citenamefont {Joly}\ \emph {et~al.}(2006)\citenamefont {Joly},
  \citenamefont {Ybert}, \citenamefont {Trizac},\ and\ \citenamefont
  {Bocquet}}]{Joly2006}%
  \BibitemOpen
  \bibfield  {author} {\bibinfo {author} {\bibfnamefont {L.}~\bibnamefont
  {Joly}}, \bibinfo {author} {\bibfnamefont {C.}~\bibnamefont {Ybert}},
  \bibinfo {author} {\bibfnamefont {E.}~\bibnamefont {Trizac}}, \ and\ \bibinfo
  {author} {\bibfnamefont {L.}~\bibnamefont {Bocquet}},\ }\href {\doibase
  10.1063/1.2397677} {\bibfield  {journal} {\bibinfo  {journal} {The Journal of
  Chemical Physics}\ }\textbf {\bibinfo {volume} {125}},\ \bibinfo {pages}
  {204716} (\bibinfo {year} {2006})}\BibitemShut {NoStop}%
\bibitem [{\citenamefont {Bouzigues}\ \emph {et~al.}(2008)\citenamefont
  {Bouzigues}, \citenamefont {Tabeling},\ and\ \citenamefont
  {Bocquet}}]{Bouzigues:2008db}%
  \BibitemOpen
  \bibfield  {author} {\bibinfo {author} {\bibfnamefont {C.~I.}\ \bibnamefont
  {Bouzigues}}, \bibinfo {author} {\bibfnamefont {P.}~\bibnamefont {Tabeling}},
  \ and\ \bibinfo {author} {\bibfnamefont {L.}~\bibnamefont {Bocquet}},\ }\href
  {\doibase 10.1103/PhysRevLett.101.114503} {\bibfield  {journal} {\bibinfo
  {journal} {Physical Review Letters}\ }\textbf {\bibinfo {volume} {101}},\
  \bibinfo {pages} {114503} (\bibinfo {year} {2008})}\BibitemShut {NoStop}%
\bibitem [{\citenamefont {Audry}\ \emph {et~al.}(2010)\citenamefont {Audry},
  \citenamefont {Piednoir}, \citenamefont {Joseph},\ and\ \citenamefont
  {Charlaix}}]{Audry:2010gy}%
  \BibitemOpen
  \bibfield  {author} {\bibinfo {author} {\bibfnamefont {M.-C.}\ \bibnamefont
  {Audry}}, \bibinfo {author} {\bibfnamefont {A.}~\bibnamefont {Piednoir}},
  \bibinfo {author} {\bibfnamefont {P.}~\bibnamefont {Joseph}}, \ and\ \bibinfo
  {author} {\bibfnamefont {E.}~\bibnamefont {Charlaix}},\ }\href {\doibase
  10.1039/B927158A} {\bibfield  {journal} {\bibinfo  {journal} {Faraday
  discussions}\ }\textbf {\bibinfo {volume} {146}},\ \bibinfo {pages}
  {113–124} (\bibinfo {year} {2010})}\BibitemShut {NoStop}%
\bibitem [{\citenamefont {Morthomas}\ and\ \citenamefont
  {W\"urger}(2009)}]{Morthomas:2009in}%
  \BibitemOpen
  \bibfield  {author} {\bibinfo {author} {\bibfnamefont {J.}~\bibnamefont
  {Morthomas}}\ and\ \bibinfo {author} {\bibfnamefont {A.}~\bibnamefont
  {W\"urger}},\ }\href {\doibase 10.1088/0953-8984/21/3/035103} {\bibfield
  {journal} {\bibinfo  {journal} {Journal of Physics: Condensed Matter}\
  }\textbf {\bibinfo {volume} {21}},\ \bibinfo {pages} {035103} (\bibinfo
  {year} {2009})}\BibitemShut {NoStop}%
\bibitem [{\citenamefont {Huang}\ \emph {et~al.}(2008)\citenamefont {Huang},
  \citenamefont {Cottin-Bizonne}, \citenamefont {Ybert},\ and\ \citenamefont
  {Bocquet}}]{Huang2008}%
  \BibitemOpen
  \bibfield  {author} {\bibinfo {author} {\bibfnamefont {D.~M.}\ \bibnamefont
  {Huang}}, \bibinfo {author} {\bibfnamefont {C.}~\bibnamefont
  {Cottin-Bizonne}}, \bibinfo {author} {\bibfnamefont {C.}~\bibnamefont
  {Ybert}}, \ and\ \bibinfo {author} {\bibfnamefont {L.}~\bibnamefont
  {Bocquet}},\ }\href {\doibase 10.1021/la7021787} {\bibfield  {journal}
  {\bibinfo  {journal} {Langmuir}\ }\textbf {\bibinfo {volume} {24}},\ \bibinfo
  {pages} {1442} (\bibinfo {year} {2008})}\BibitemShut {NoStop}%
\bibitem [{\citenamefont {Barbosa De~Lima}\ and\ \citenamefont
  {Joly}(2017)}]{BarbosaDeLima2017}%
  \BibitemOpen
  \bibfield  {author} {\bibinfo {author} {\bibfnamefont {A.}~\bibnamefont
  {Barbosa De~Lima}}\ and\ \bibinfo {author} {\bibfnamefont {L.}~\bibnamefont
  {Joly}},\ }\href {\doibase 10.1039/C7SM00358G} {\bibfield  {journal}
  {\bibinfo  {journal} {Soft Matter}\ }\textbf {\bibinfo {volume} {13}},\
  \bibinfo {pages} {3341} (\bibinfo {year} {2017})}\BibitemShut {NoStop}%
\bibitem [{\citenamefont {Plimpton}(1995)}]{Plimpton1995}%
  \BibitemOpen
  \bibfield  {author} {\bibinfo {author} {\bibfnamefont {S.}~\bibnamefont
  {Plimpton}},\ }\href {\doibase 10.1006/jcph.1995.1039} {\bibfield  {journal}
  {\bibinfo  {journal} {Journal of Computational Physics}\ }\textbf {\bibinfo
  {volume} {117}},\ \bibinfo {pages} {1} (\bibinfo {year} {1995})}\BibitemShut
  {NoStop}%
\bibitem [{\citenamefont {Derjaguin}(1980)}]{Derjaguin1980}%
  \BibitemOpen
  \bibfield  {author} {\bibinfo {author} {\bibfnamefont {B.~V.}\ \bibnamefont
  {Derjaguin}},\ }\href {\doibase 10.1351/pac198052051163} {\bibfield
  {journal} {\bibinfo  {journal} {Pure and Applied Chemistry}\ }\textbf
  {\bibinfo {volume} {52}},\ \bibinfo {pages} {1163} (\bibinfo {year}
  {1980})}\BibitemShut {NoStop}%
\bibitem [{\citenamefont {Rusconi}\ \emph {et~al.}(2004)\citenamefont
  {Rusconi}, \citenamefont {Isa},\ and\ \citenamefont {Piazza}}]{Rusconi2004}%
  \BibitemOpen
  \bibfield  {author} {\bibinfo {author} {\bibfnamefont {R.}~\bibnamefont
  {Rusconi}}, \bibinfo {author} {\bibfnamefont {L.}~\bibnamefont {Isa}}, \ and\
  \bibinfo {author} {\bibfnamefont {R.}~\bibnamefont {Piazza}},\ }\href
  {\doibase 10.1364/JOSAB.21.000605} {\bibfield  {journal} {\bibinfo  {journal}
  {Journal of the Optical Society of America B}\ }\textbf {\bibinfo {volume}
  {21}},\ \bibinfo {pages} {605} (\bibinfo {year} {2004})}\BibitemShut
  {NoStop}%
\bibitem [{\citenamefont {L{\"{u}}sebrink}\ \emph {et~al.}(2012)\citenamefont
  {L{\"{u}}sebrink}, \citenamefont {Yang},\ and\ \citenamefont
  {Ripoll}}]{Lusebrink2012}%
  \BibitemOpen
  \bibfield  {author} {\bibinfo {author} {\bibfnamefont {D.}~\bibnamefont
  {L{\"{u}}sebrink}}, \bibinfo {author} {\bibfnamefont {M.}~\bibnamefont
  {Yang}}, \ and\ \bibinfo {author} {\bibfnamefont {M.}~\bibnamefont
  {Ripoll}},\ }\href {\doibase 10.1088/0953-8984/24/28/284132} {\bibfield
  {journal} {\bibinfo  {journal} {Journal of Physics: Condensed Matter}\
  }\textbf {\bibinfo {volume} {24}},\ \bibinfo {pages} {284132} (\bibinfo
  {year} {2012})}\BibitemShut {NoStop}%
\bibitem [{\citenamefont {Nedev}\ \emph {et~al.}(2015)\citenamefont {Nedev},
  \citenamefont {Carretero-Palacios}, \citenamefont {K{\"{u}}hler},
  \citenamefont {Lohm{\"{u}}ller}, \citenamefont {Urban}, \citenamefont
  {Anderson},\ and\ \citenamefont {Feldmann}}]{Nedev2015}%
  \BibitemOpen
  \bibfield  {author} {\bibinfo {author} {\bibfnamefont {S.}~\bibnamefont
  {Nedev}}, \bibinfo {author} {\bibfnamefont {S.}~\bibnamefont
  {Carretero-Palacios}}, \bibinfo {author} {\bibfnamefont {P.}~\bibnamefont
  {K{\"{u}}hler}}, \bibinfo {author} {\bibfnamefont {T.}~\bibnamefont
  {Lohm{\"{u}}ller}}, \bibinfo {author} {\bibfnamefont {A.~S.}\ \bibnamefont
  {Urban}}, \bibinfo {author} {\bibfnamefont {L.~J.~E.}\ \bibnamefont
  {Anderson}}, \ and\ \bibinfo {author} {\bibfnamefont {J.}~\bibnamefont
  {Feldmann}},\ }\href {\doibase 10.1021/ph500371z} {\bibfield  {journal}
  {\bibinfo  {journal} {ACS Photonics}\ }\textbf {\bibinfo {volume} {2}},\
  \bibinfo {pages} {491} (\bibinfo {year} {2015})}\BibitemShut {NoStop}%
\bibitem [{\citenamefont {Ma}\ and\ \citenamefont {Hill}(2006)}]{Ma2006}%
  \BibitemOpen
  \bibfield  {author} {\bibinfo {author} {\bibfnamefont {M.}~\bibnamefont
  {Ma}}\ and\ \bibinfo {author} {\bibfnamefont {R.~M.}\ \bibnamefont {Hill}},\
  }\href {\doibase 10.1016/j.cocis.2006.06.002} {\bibfield  {journal} {\bibinfo
   {journal} {Current Opinion in Colloid {\&} Interface Science}\ }\textbf
  {\bibinfo {volume} {11}},\ \bibinfo {pages} {193} (\bibinfo {year}
  {2006})}\BibitemShut {NoStop}%
\bibitem [{\citenamefont {Ybert}\ \emph {et~al.}(2007)\citenamefont {Ybert},
  \citenamefont {Barentin}, \citenamefont {Cottin-Bizonne}, \citenamefont
  {Joseph},\ and\ \citenamefont {Bocquet}}]{Ybert2007}%
  \BibitemOpen
  \bibfield  {author} {\bibinfo {author} {\bibfnamefont {C.}~\bibnamefont
  {Ybert}}, \bibinfo {author} {\bibfnamefont {C.}~\bibnamefont {Barentin}},
  \bibinfo {author} {\bibfnamefont {C.}~\bibnamefont {Cottin-Bizonne}},
  \bibinfo {author} {\bibfnamefont {P.}~\bibnamefont {Joseph}}, \ and\ \bibinfo
  {author} {\bibfnamefont {L.}~\bibnamefont {Bocquet}},\ }\href {\doibase
  10.1063/1.2815730} {\bibfield  {journal} {\bibinfo  {journal} {Physics of
  Fluids}\ }\textbf {\bibinfo {volume} {19}},\ \bibinfo {pages} {123601}
  (\bibinfo {year} {2007})}\BibitemShut {NoStop}%
\bibitem [{\citenamefont {Baier}\ \emph {et~al.}(2010)\citenamefont {Baier},
  \citenamefont {Steffes},\ and\ \citenamefont {Hardt}}]{Baier2010}%
  \BibitemOpen
  \bibfield  {author} {\bibinfo {author} {\bibfnamefont {T.}~\bibnamefont
  {Baier}}, \bibinfo {author} {\bibfnamefont {C.}~\bibnamefont {Steffes}}, \
  and\ \bibinfo {author} {\bibfnamefont {S.}~\bibnamefont {Hardt}},\ }\href
  {\doibase 10.1103/PhysRevE.82.037301} {\bibfield  {journal} {\bibinfo
  {journal} {Physical Review E}\ }\textbf {\bibinfo {volume} {82}},\ \bibinfo
  {pages} {037301} (\bibinfo {year} {2010})}\BibitemShut {NoStop}%
\bibitem [{\citenamefont {Sampson}(1891)}]{Sampson:1891hj}%
  \BibitemOpen
  \bibfield  {author} {\bibinfo {author} {\bibfnamefont {R.~A.}\ \bibnamefont
  {Sampson}},\ }\href {\doibase 10.1098/rsta.1891.0012} {\bibfield  {journal}
  {\bibinfo  {journal} {Philosophical Transactions of the Royal Society of
  London A: Mathematical, Physical and Engineering Sciences}\ }\textbf
  {\bibinfo {volume} {182}},\ \bibinfo {pages} {449–518} (\bibinfo {year}
  {1891})}\BibitemShut {NoStop}%
\bibitem [{\citenamefont {Weissberg}(1962)}]{Weissberg:1962iz}%
  \BibitemOpen
  \bibfield  {author} {\bibinfo {author} {\bibfnamefont {H.~L.}\ \bibnamefont
  {Weissberg}},\ }\href {\doibase 10.1063/1.1724469} {\bibfield  {journal}
  {\bibinfo  {journal} {Physics of Fluids}\ }\textbf {\bibinfo {volume} {5}},\
  \bibinfo {pages} {1033} (\bibinfo {year} {1962})}\BibitemShut {NoStop}%
\bibitem [{\citenamefont {Dagan}\ and\ \citenamefont
  {Weinbaum}(1982)}]{Dagan:1982wu}%
  \BibitemOpen
  \bibfield  {author} {\bibinfo {author} {\bibfnamefont {Z.}~\bibnamefont
  {Dagan}}\ and\ \bibinfo {author} {\bibfnamefont {S.}~\bibnamefont
  {Weinbaum}},\ }\href@noop {} {\bibfield  {journal} {\bibinfo  {journal}
  {Journal of Fluid …}\ } (\bibinfo {year} {1982})}\BibitemShut {NoStop}%
\bibitem [{\citenamefont {Sisan}\ and\ \citenamefont
  {Lichter}(2011)}]{Sisan2011}%
  \BibitemOpen
  \bibfield  {author} {\bibinfo {author} {\bibfnamefont {T.~B.}\ \bibnamefont
  {Sisan}}\ and\ \bibinfo {author} {\bibfnamefont {S.}~\bibnamefont
  {Lichter}},\ }\href {\doibase 10.1007/s10404-011-0855-9} {\bibfield
  {journal} {\bibinfo  {journal} {Microfluidics and Nanofluidics}\ }\textbf
  {\bibinfo {volume} {11}},\ \bibinfo {pages} {787} (\bibinfo {year}
  {2011})}\BibitemShut {NoStop}%
\bibitem [{\citenamefont {Joly}(2011)}]{Joly2011}%
  \BibitemOpen
  \bibfield  {author} {\bibinfo {author} {\bibfnamefont {L.}~\bibnamefont
  {Joly}},\ }\href {\doibase 10.1063/1.3664622} {\bibfield  {journal} {\bibinfo
   {journal} {The Journal of Chemical Physics}\ }\textbf {\bibinfo {volume}
  {135}},\ \bibinfo {pages} {214705} (\bibinfo {year} {2011})}\BibitemShut
  {NoStop}%
\bibitem [{\citenamefont {Gravelle}\ \emph {et~al.}(2013)\citenamefont
  {Gravelle}, \citenamefont {Joly}, \citenamefont {Detcheverry}, \citenamefont
  {Ybert}, \citenamefont {Cottin-Bizonne},\ and\ \citenamefont
  {Bocquet}}]{Gravelle2013}%
  \BibitemOpen
  \bibfield  {author} {\bibinfo {author} {\bibfnamefont {S.}~\bibnamefont
  {Gravelle}}, \bibinfo {author} {\bibfnamefont {L.}~\bibnamefont {Joly}},
  \bibinfo {author} {\bibfnamefont {F.}~\bibnamefont {Detcheverry}}, \bibinfo
  {author} {\bibfnamefont {C.}~\bibnamefont {Ybert}}, \bibinfo {author}
  {\bibfnamefont {C.}~\bibnamefont {Cottin-Bizonne}}, \ and\ \bibinfo {author}
  {\bibfnamefont {L.}~\bibnamefont {Bocquet}},\ }\href {\doibase
  10.1073/pnas.1306447110} {\bibfield  {journal} {\bibinfo  {journal} {Proc.
  Natl. Acad. Sci. U. S. A.}\ }\textbf {\bibinfo {volume} {110}},\ \bibinfo
  {pages} {16367} (\bibinfo {year} {2013})}\BibitemShut {NoStop}%
\bibitem [{\citenamefont {Walther}\ \emph {et~al.}(2013)\citenamefont
  {Walther}, \citenamefont {Ritos}, \citenamefont {Cruz}, \citenamefont
  {Megaridis}, \citenamefont {Koumoutsakos},\ and\ \citenamefont
  {Cruz-chu}}]{Walther2013}%
  \BibitemOpen
  \bibfield  {author} {\bibinfo {author} {\bibfnamefont {J.~H.}\ \bibnamefont
  {Walther}}, \bibinfo {author} {\bibfnamefont {K.}~\bibnamefont {Ritos}},
  \bibinfo {author} {\bibfnamefont {E.}~\bibnamefont {Cruz}}, \bibinfo {author}
  {\bibfnamefont {C.~M.}\ \bibnamefont {Megaridis}}, \bibinfo {author}
  {\bibfnamefont {P.}~\bibnamefont {Koumoutsakos}}, \ and\ \bibinfo {author}
  {\bibfnamefont {E.~R.}\ \bibnamefont {Cruz-chu}},\ }\href {\doibase
  10.1021/nl304000k} {\bibfield  {journal} {\bibinfo  {journal} {Nano Letters}\
  }\textbf {\bibinfo {volume} {13}},\ \bibinfo {pages} {1910–1914} (\bibinfo
  {year} {2013})}\BibitemShut {NoStop}%
\bibitem [{\citenamefont {Gravelle}\ \emph {et~al.}(2014)\citenamefont
  {Gravelle}, \citenamefont {Joly}, \citenamefont {Ybert},\ and\ \citenamefont
  {Bocquet}}]{Gravelle:2014jj}%
  \BibitemOpen
  \bibfield  {author} {\bibinfo {author} {\bibfnamefont {S.}~\bibnamefont
  {Gravelle}}, \bibinfo {author} {\bibfnamefont {L.}~\bibnamefont {Joly}},
  \bibinfo {author} {\bibfnamefont {C.}~\bibnamefont {Ybert}}, \ and\ \bibinfo
  {author} {\bibfnamefont {L.}~\bibnamefont {Bocquet}},\ }\href {\doibase
  10.1063/1.4897253} {\bibfield  {journal} {\bibinfo  {journal} {The Journal of
  Chemical Physics}\ }\textbf {\bibinfo {volume} {141}},\ \bibinfo {pages}
  {18C526} (\bibinfo {year} {2014})}\BibitemShut {NoStop}%
\bibitem [{\citenamefont {Abascal}\ and\ \citenamefont
  {Vega}(2005)}]{Abascal:2005ka}%
  \BibitemOpen
  \bibfield  {author} {\bibinfo {author} {\bibfnamefont {J.~L.~F.}\
  \bibnamefont {Abascal}}\ and\ \bibinfo {author} {\bibfnamefont
  {C.}~\bibnamefont {Vega}},\ }\href {\doibase 10.1063/1.2121687} {\bibfield
  {journal} {\bibinfo  {journal} {The Journal of Chemical Physics}\ }\textbf
  {\bibinfo {volume} {123}},\ \bibinfo {pages} {234505} (\bibinfo {year}
  {2005})}\BibitemShut {NoStop}%
\bibitem [{\citenamefont {Los}\ and\ \citenamefont
  {Fasolino}(2003)}]{Los:2003hd}%
  \BibitemOpen
  \bibfield  {author} {\bibinfo {author} {\bibfnamefont {J.~H.}\ \bibnamefont
  {Los}}\ and\ \bibinfo {author} {\bibfnamefont {A.}~\bibnamefont {Fasolino}},\
  }\href {\doibase 10.1103/PhysRevB.68.024107} {\bibfield  {journal} {\bibinfo
  {journal} {Physical Review B}\ }\textbf {\bibinfo {volume} {68}},\ \bibinfo
  {pages} {024107} (\bibinfo {year} {2003})}\BibitemShut {NoStop}%
\bibitem [{\citenamefont {P\'erez-Hern\'andez}\ and\ \citenamefont
  {Schmidt}(2013)}]{PerezHernandez:2013dr}%
  \BibitemOpen
  \bibfield  {author} {\bibinfo {author} {\bibfnamefont {G.}~\bibnamefont
  {P\'erez-Hern\'andez}}\ and\ \bibinfo {author} {\bibfnamefont
  {B.}~\bibnamefont {Schmidt}},\ }\href {\doibase 10.1039/C3CP44278K}
  {\bibfield  {journal} {\bibinfo  {journal} {Physical Chemistry Chemical
  Physics}\ }\textbf {\bibinfo {volume} {15}},\ \bibinfo {pages} {4995}
  (\bibinfo {year} {2013})}\BibitemShut {NoStop}%
\bibitem [{\citenamefont {Al-Hamdani}\ \emph {et~al.}(2017)\citenamefont
  {Al-Hamdani}, \citenamefont {Alf{\`{e}}},\ and\ \citenamefont
  {Michaelides}}]{Al-Hamdani2017}%
  \BibitemOpen
  \bibfield  {author} {\bibinfo {author} {\bibfnamefont {Y.~S.}\ \bibnamefont
  {Al-Hamdani}}, \bibinfo {author} {\bibfnamefont {D.}~\bibnamefont
  {Alf{\`{e}}}}, \ and\ \bibinfo {author} {\bibfnamefont {A.}~\bibnamefont
  {Michaelides}},\ }\href {\doibase 10.1063/1.4977180} {\bibfield  {journal}
  {\bibinfo  {journal} {The Journal of Chemical Physics}\ }\textbf {\bibinfo
  {volume} {146}},\ \bibinfo {pages} {094701} (\bibinfo {year}
  {2017})}\BibitemShut {NoStop}%
\bibitem [{\citenamefont {Wang}\ \emph {et~al.}(2016)\citenamefont {Wang},
  \citenamefont {Zhang}, \citenamefont {Li},\ and\ \citenamefont
  {Zhan}}]{Wang2016}%
  \BibitemOpen
  \bibfield  {author} {\bibinfo {author} {\bibfnamefont {W.}~\bibnamefont
  {Wang}}, \bibinfo {author} {\bibfnamefont {H.}~\bibnamefont {Zhang}},
  \bibinfo {author} {\bibfnamefont {S.}~\bibnamefont {Li}}, \ and\ \bibinfo
  {author} {\bibfnamefont {Y.}~\bibnamefont {Zhan}},\ }\href {\doibase
  10.1088/0957-4484/27/7/075707} {\bibfield  {journal} {\bibinfo  {journal}
  {Nanotechnology}\ }\textbf {\bibinfo {volume} {27}},\ \bibinfo {pages}
  {075707} (\bibinfo {year} {2016})}\BibitemShut {NoStop}%
\bibitem [{\citenamefont {Kannam}\ \emph {et~al.}(2013)\citenamefont {Kannam},
  \citenamefont {Todd}, \citenamefont {Hansen},\ and\ \citenamefont
  {Daivis}}]{Kannam:2013gf}%
  \BibitemOpen
  \bibfield  {author} {\bibinfo {author} {\bibfnamefont {S.~K.}\ \bibnamefont
  {Kannam}}, \bibinfo {author} {\bibfnamefont {B.~D.}\ \bibnamefont {Todd}},
  \bibinfo {author} {\bibfnamefont {J.~S.}\ \bibnamefont {Hansen}}, \ and\
  \bibinfo {author} {\bibfnamefont {P.~J.}\ \bibnamefont {Daivis}},\ }\href
  {\doibase 10.1063/1.4793396} {\bibfield  {journal} {\bibinfo  {journal} {The
  Journal of Chemical Physics}\ }\textbf {\bibinfo {volume} {138}},\ \bibinfo
  {pages} {094701} (\bibinfo {year} {2013})}\BibitemShut {NoStop}%
\bibitem [{\citenamefont {Bocquet}\ and\ \citenamefont
  {Barrat}(2007)}]{Bocquet2007}%
  \BibitemOpen
  \bibfield  {author} {\bibinfo {author} {\bibfnamefont {L.}~\bibnamefont
  {Bocquet}}\ and\ \bibinfo {author} {\bibfnamefont {J.-L.}\ \bibnamefont
  {Barrat}},\ }\href {\doibase 10.1039/b616490k} {\bibfield  {journal}
  {\bibinfo  {journal} {Soft Matter}\ }\textbf {\bibinfo {volume} {3}},\
  \bibinfo {pages} {685} (\bibinfo {year} {2007})}\BibitemShut {NoStop}%
\bibitem [{\citenamefont {Balandin}\ \emph {et~al.}(2008)\citenamefont
  {Balandin}, \citenamefont {Ghosh}, \citenamefont {Bao}, \citenamefont
  {Calizo}, \citenamefont {Teweldebrhan}, \citenamefont {Miao},\ and\
  \citenamefont {Lau}}]{Balandin2008}%
  \BibitemOpen
  \bibfield  {author} {\bibinfo {author} {\bibfnamefont {A.~A.}\ \bibnamefont
  {Balandin}}, \bibinfo {author} {\bibfnamefont {S.}~\bibnamefont {Ghosh}},
  \bibinfo {author} {\bibfnamefont {W.}~\bibnamefont {Bao}}, \bibinfo {author}
  {\bibfnamefont {I.}~\bibnamefont {Calizo}}, \bibinfo {author} {\bibfnamefont
  {D.}~\bibnamefont {Teweldebrhan}}, \bibinfo {author} {\bibfnamefont
  {F.}~\bibnamefont {Miao}}, \ and\ \bibinfo {author} {\bibfnamefont {C.~N.}\
  \bibnamefont {Lau}},\ }\href {\doibase 10.1021/nl0731872} {\bibfield
  {journal} {\bibinfo  {journal} {Nano Letters}\ }\textbf {\bibinfo {volume}
  {8}},\ \bibinfo {pages} {902} (\bibinfo {year} {2008})}\BibitemShut {NoStop}%
\bibitem [{\citenamefont {Jewett}\ \emph {et~al.}(2013)\citenamefont {Jewett},
  \citenamefont {Zhuang},\ and\ \citenamefont {Shea}}]{Jewett:2013fg}%
  \BibitemOpen
  \bibfield  {author} {\bibinfo {author} {\bibfnamefont {A.~I.}\ \bibnamefont
  {Jewett}}, \bibinfo {author} {\bibfnamefont {Z.}~\bibnamefont {Zhuang}}, \
  and\ \bibinfo {author} {\bibfnamefont {J.-E.}\ \bibnamefont {Shea}},\ }\href
  {\doibase 10.1016/j.bpj.2012.11.953} {\bibfield  {journal} {\bibinfo
  {journal} {Biophys J}\ }\textbf {\bibinfo {volume} {104}},\ \bibinfo {pages}
  {169a} (\bibinfo {year} {2013})}\BibitemShut {NoStop}%
\bibitem [{\citenamefont {Humphrey}\ \emph {et~al.}(1996)\citenamefont
  {Humphrey}, \citenamefont {Dalke},\ and\ \citenamefont
  {Schulten}}]{Humphrey:1996bk}%
  \BibitemOpen
  \bibfield  {author} {\bibinfo {author} {\bibfnamefont {W.}~\bibnamefont
  {Humphrey}}, \bibinfo {author} {\bibfnamefont {A.}~\bibnamefont {Dalke}}, \
  and\ \bibinfo {author} {\bibfnamefont {K.}~\bibnamefont {Schulten}},\ }\href
  {\doibase 10.1016/0263-7855(96)00018-5} {\bibfield  {journal} {\bibinfo
  {journal} {Journal of molecular graphics}\ }\textbf {\bibinfo {volume}
  {14}},\ \bibinfo {pages} {33–38} (\bibinfo {year} {1996})}\BibitemShut
  {NoStop}%
\bibitem [{\citenamefont {Bussi}\ \emph {et~al.}(2007)\citenamefont {Bussi},
  \citenamefont {Donadio},\ and\ \citenamefont {Parrinello}}]{Bussi2007}%
  \BibitemOpen
  \bibfield  {author} {\bibinfo {author} {\bibfnamefont {G.}~\bibnamefont
  {Bussi}}, \bibinfo {author} {\bibfnamefont {D.}~\bibnamefont {Donadio}}, \
  and\ \bibinfo {author} {\bibfnamefont {M.}~\bibnamefont {Parrinello}},\
  }\href {\doibase 10.1063/1.2408420} {\bibfield  {journal} {\bibinfo
  {journal} {The Journal of Chemical Physics}\ }\textbf {\bibinfo {volume}
  {126}},\ \bibinfo {pages} {014101} (\bibinfo {year} {2007})}\BibitemShut
  {NoStop}%
\bibitem [{\citenamefont {Barrat}\ and\ \citenamefont
  {Bocquet}(1999)}]{Barrat:1999kv}%
  \BibitemOpen
  \bibfield  {author} {\bibinfo {author} {\bibfnamefont {J.-L.}\ \bibnamefont
  {Barrat}}\ and\ \bibinfo {author} {\bibfnamefont {L.}~\bibnamefont
  {Bocquet}},\ }\href {\doibase 10.1103/PhysRevLett.82.4671} {\bibfield
  {journal} {\bibinfo  {journal} {Physical Review Letters}\ }\textbf {\bibinfo
  {volume} {82}},\ \bibinfo {pages} {4671–4674} (\bibinfo {year}
  {1999})}\BibitemShut {NoStop}%
\bibitem [{\citenamefont {Evans}\ \emph {et~al.}(2016)\citenamefont {Evans},
  \citenamefont {Stewart},\ and\ \citenamefont {Wilding}}]{Evans:2016ur}%
  \BibitemOpen
  \bibfield  {author} {\bibinfo {author} {\bibfnamefont {R.}~\bibnamefont
  {Evans}}, \bibinfo {author} {\bibfnamefont {M.~C.}\ \bibnamefont {Stewart}},
  \ and\ \bibinfo {author} {\bibfnamefont {N.~B.}\ \bibnamefont {Wilding}},\
  }\href {\doibase 10.1103/PhysRevLett.117.176102} {\bibfield  {journal}
  {\bibinfo  {journal} {Phys. Rev. Lett.}\ }\textbf {\bibinfo {volume} {117}},\
  \bibinfo {pages} {176102} (\bibinfo {year} {2016})}\BibitemShut {NoStop}%
\bibitem [{\citenamefont {Tocci}\ \emph {et~al.}(2014)\citenamefont {Tocci},
  \citenamefont {Joly},\ and\ \citenamefont {Michaelides}}]{Tocci2014a}%
  \BibitemOpen
  \bibfield  {author} {\bibinfo {author} {\bibfnamefont {G.}~\bibnamefont
  {Tocci}}, \bibinfo {author} {\bibfnamefont {L.}~\bibnamefont {Joly}}, \ and\
  \bibinfo {author} {\bibfnamefont {A.}~\bibnamefont {Michaelides}},\ }\href
  {\doibase 10.1021/nl502837d} {\bibfield  {journal} {\bibinfo  {journal} {Nano
  Lett.}\ }\textbf {\bibinfo {volume} {14}},\ \bibinfo {pages} {6872} (\bibinfo
  {year} {2014})}\BibitemShut {NoStop}%
\bibitem [{\citenamefont {Falk}\ \emph {et~al.}(2010)\citenamefont {Falk},
  \citenamefont {Sedlmeier}, \citenamefont {Joly}, \citenamefont {Netz},\ and\
  \citenamefont {Bocquet}}]{Falk2010}%
  \BibitemOpen
  \bibfield  {author} {\bibinfo {author} {\bibfnamefont {K.}~\bibnamefont
  {Falk}}, \bibinfo {author} {\bibfnamefont {F.}~\bibnamefont {Sedlmeier}},
  \bibinfo {author} {\bibfnamefont {L.}~\bibnamefont {Joly}}, \bibinfo {author}
  {\bibfnamefont {R.~R.}\ \bibnamefont {Netz}}, \ and\ \bibinfo {author}
  {\bibfnamefont {L.}~\bibnamefont {Bocquet}},\ }\href {\doibase
  10.1021/nl1021046} {\bibfield  {journal} {\bibinfo  {journal} {Nano letters}\
  }\textbf {\bibinfo {volume} {10}},\ \bibinfo {pages} {4067} (\bibinfo {year}
  {2010})}\BibitemShut {NoStop}%
\bibitem [{\citenamefont {Kannam}\ \emph {et~al.}(2012)\citenamefont {Kannam},
  \citenamefont {Todd}, \citenamefont {Hansen},\ and\ \citenamefont
  {Daivis}}]{KumarKannam2012}%
  \BibitemOpen
  \bibfield  {author} {\bibinfo {author} {\bibfnamefont {S.~K.}\ \bibnamefont
  {Kannam}}, \bibinfo {author} {\bibfnamefont {B.~D.}\ \bibnamefont {Todd}},
  \bibinfo {author} {\bibfnamefont {J.~S.}\ \bibnamefont {Hansen}}, \ and\
  \bibinfo {author} {\bibfnamefont {P.~J.}\ \bibnamefont {Daivis}},\ }\href
  {\doibase 10.1063/1.3675904} {\bibfield  {journal} {\bibinfo  {journal} {The
  Journal of Chemical Physics}\ }\textbf {\bibinfo {volume} {136}},\ \bibinfo
  {pages} {024705} (\bibinfo {year} {2012})}\BibitemShut {NoStop}%
\bibitem [{\citenamefont {Guevara-Carrion}\ \emph {et~al.}(2011)\citenamefont
  {Guevara-Carrion}, \citenamefont {Vrabec},\ and\ \citenamefont
  {Hasse}}]{Guevara-Carrion2011}%
  \BibitemOpen
  \bibfield  {author} {\bibinfo {author} {\bibfnamefont {G.}~\bibnamefont
  {Guevara-Carrion}}, \bibinfo {author} {\bibfnamefont {J.}~\bibnamefont
  {Vrabec}}, \ and\ \bibinfo {author} {\bibfnamefont {H.}~\bibnamefont
  {Hasse}},\ }\href {\doibase 10.1063/1.3515262} {\bibfield  {journal}
  {\bibinfo  {journal} {The Journal of Chemical Physics}\ }\textbf {\bibinfo
  {volume} {134}},\ \bibinfo {pages} {074508} (\bibinfo {year}
  {2011})}\BibitemShut {NoStop}%
\bibitem [{\citenamefont {Guillaud}\ \emph {et~al.}(2017)\citenamefont
  {Guillaud}, \citenamefont {Merabia}, \citenamefont {de~Ligny},\ and\
  \citenamefont {Joly}}]{Guillaud:2017bk}%
  \BibitemOpen
  \bibfield  {author} {\bibinfo {author} {\bibfnamefont {E.}~\bibnamefont
  {Guillaud}}, \bibinfo {author} {\bibfnamefont {S.}~\bibnamefont {Merabia}},
  \bibinfo {author} {\bibfnamefont {D.}~\bibnamefont {de~Ligny}}, \ and\
  \bibinfo {author} {\bibfnamefont {L.}~\bibnamefont {Joly}},\ }\href@noop {}
  {\bibfield  {journal} {\bibinfo  {journal} {Physical Chemistry Chemical
  Physics}\ }\textbf {\bibinfo {volume} {19}},\ \bibinfo {pages} {2124}
  (\bibinfo {year} {2017})}\BibitemShut {NoStop}%
\bibitem [{\citenamefont {Markesteijn}\ \emph {et~al.}(2012)\citenamefont
  {Markesteijn}, \citenamefont {Hartkamp}, \citenamefont {Luding},\ and\
  \citenamefont {Westerweel}}]{Markesteijn2012}%
  \BibitemOpen
  \bibfield  {author} {\bibinfo {author} {\bibfnamefont {A.~P.}\ \bibnamefont
  {Markesteijn}}, \bibinfo {author} {\bibfnamefont {R.}~\bibnamefont
  {Hartkamp}}, \bibinfo {author} {\bibfnamefont {S.}~\bibnamefont {Luding}}, \
  and\ \bibinfo {author} {\bibfnamefont {J.}~\bibnamefont {Westerweel}},\
  }\href {\doibase 10.1063/1.3697977} {\bibfield  {journal} {\bibinfo
  {journal} {The Journal of Chemical Physics}\ }\textbf {\bibinfo {volume}
  {136}},\ \bibinfo {pages} {134104} (\bibinfo {year} {2012})}\BibitemShut
  {NoStop}%
\end{thebibliography}%

\end{document}